\documentclass[a4paper,preprintnumbers,amsmath,amssymb,nofootinbib,showpacs,10pt]{revtex4}
\usepackage[margin=2cm]{geometry}
\usepackage{amssymb}
\usepackage{mathrsfs}
\usepackage{graphics}

\ifx\pdfoutput\undefined
\usepackage{graphicx}
\else
\usepackage[pdftex]{graphicx}
\fi

\def\P{{\cal{P}}}

\def\be{\begin{equation}}
\def\ee{\end{equation}}
\def\bea{\begin{eqnarray}}
\def\eea{\end{eqnarray}}

\def\bx{{\bf x}}

\def\bq{\mathbf q}

\def\bqone{\mathbf q_1}
\def\bqtwo{\mathbf q_2}

\def\bkone{\mathbf k_1}
\def\bktwo{\mathbf k_2}
\def\bkthree{\mathbf k_3}

\newcommand\lsim{\mathrel{\rlap{\lower4pt\hbox{\hskip1pt$\sim$}}
    \raise1pt\hbox{$<$}}}
\newcommand\gsim{\mathrel{\rlap{\lower4pt\hbox{\hskip1pt$\sim$}}
    \raise1pt\hbox{$>$}}}

\begin{document}

\title{On the Physical Significance of Infra-red \\ Corrections to 
Inflationary Observables}
\author{N. Bartolo$^{1,2}$,
S. Matarrese$^{1,2}$, 
M. Pietroni$^{2}$,
A. Riotto$^{2,3}$ and
D. Seery$^4$}

\affiliation{$^1$ Dipartimento di Fisica, Universit\'a di  Padova, Via Marzolo,
8 - I-35131 Padua -- Italy\\}

\affiliation{$^2$ INFN, Sezione di Padova, Via Marzolo,
8 - I-35131 Padua -- Italy\\}

\affiliation{$^3$ CERN, Theory Division, Gen\'eve 23, CH-1211 Switzerland\\}
% \date{\today}

\affiliation{$^4$ Centre for Theoretical Cosmology,
Department of Applied Mathematics and Theoretical Physics,
University of Cambridge,
Wilberforce Road, Cambridge, CB3 0WA -- United Kingdom\\}

\pacs{98.80.Cq  \hfill CERN-PH-TH/2007-214}

\begin{abstract}
\noindent
Inflationary observables, like the power 
spectrum, computed at one- and higher-order loop level seem 
to be plagued by large infra-red corrections. 
In this short note, we point out  that these large infra-red corrections 
appear only in quantities which are not directly observable.
This is in agreement with general expectations concerning infra-red effects. 
\end{abstract}

\maketitle
\noindent
Cosmological inflation \cite{lrreview} has become the dominant paradigm
within which one can attempt to
understand the initial conditions for Cosmic Microwave Background (CMB)
anisotropies and structure formation. In the inflationary picture, 
the primordial cosmological perturbations are created from quantum 
fluctuations which are
``redshifted'' out of the horizon during an early period of 
accelerated expansion.
Once outside the horizon, they remain ``frozen'' until the horizon
grows during a later matter- or radiation-dominated era.
After falling back inside the horizon they are communicated to the primordial
plasma and hence are directly
observable as temperature anisotropies in the CMB.
These anisotropies were first detected by the Cosmic Background 
Explorer (COBE) satellite \cite{smoot92,bennett96,gorski96},
and more recently they have been mapped 
with spectacular accuracy by the  Wilkinson Microwave 
Anisotropy Probe (WMAP)~\cite{wmap}.

These CMB observations show that the cosmological perturbations are very small,
of order $10^{-5}$ compared to the homogeneous background.
Therefore, one might think that first-order perturbation theory
will be adequate for all comparison with observations.
However, that may not be
the case; the Planck satellite \cite{planck} and its
successors may be sensitive to non-Gaussianity (NG) in the cosmological
perturbations.
Such non-Gaussianities are sourced by self-interactions in the early universe,
and become visible
at the level of second- or higher-order perturbation theory \cite{review}.
This possibility is of considerable interest in its own right
and is presently being vigorously explored.
However, there are other equally compelling reasons to go beyond linear
theory.
Self-interactions of any scalar field during the inflationary stage, and
more interestingly of the comoving curvature perturbation
$\zeta$, do not \emph{only} imply non-Gaussian statistics:
they also give rise to  corrections in all correlators,
including the observationally interesting cases of the
two- and the three-point correlation functions.
Such  corrections are associated with so-called ``loops,''
in which virtual particles with arbitrary momentum are emitted and
re-absorbed by the fields which participate in the correlation function. 

Loop corrections may lead to significant effects
\cite{Mukhanov:1996ak,Abramo:1997hu,boy1,boy2,olandesi}
which scale like (powers of)  
the number of e-folds between horizon exit of
the mode $k$ under consideration and the end of inflation.
Furthermore, it has been known for a long time that there exists a breakdown
in the perturbative
expansion due to infra-red (IR) divergences \cite{sasaki1,sasaki2}. 
These  classical one-loop corrections depend on the total 
number of e-folds of inflation since they scale
like powers of $\ln a$, where $a$ is the scale factor of the universe 
\cite{Sloth:2006az,Sloth:2006nu,Weinberg:2005vy,wein2,seery1,seery2}.

To illustrate the appearance of IR divergences, we can adopt 
a simple toy model made of a scalar field $\phi$ with cubic interaction
$\mu\phi^3/3!$. This field is not necessarily the scalar field
driving inflation. By using the in--in formalism suitable for non-equilibrium
field theory, we can compute the one-loop correction to the power spectrum
$\P_\phi$
of the scalar field perturbations in a pure de Sitter epoch characterized
by a constant Hubble rate $H$. The computation is performed
in Appendix A, to which the reader is referred for all technical details.
One finds
\begin{equation}
	\P^{\rm 1-loop}_\phi=
	\frac{2\mu^2}{9(2\pi)^2H^2}\P^{\rm tree}_\phi
	\ln(k L)\ln^2(k\tau),\,\,\,\
\P^{\rm tree}_\phi=\left(\frac{H}{2\pi}\right)^2,
\end{equation}
where $\tau$ is the conformal time and 
$L^{-1}$ is the infra-red comoving momentum cut-off.
One is now faced with the task of determining what value should be assigned
to $L^{-1}$.

One option is 
to set $L$ to be some comoving length scale which left the
horizon many e-folds before the observable universe. In particular,
the smallest possible value of $L^{-1}$ is  
$a_i H$, where $a_i$ is the value of the scale factor at the beginning
of inflation. Since the wavelength
$k^{-1}$ goes out of the horizon when the scale factor equals
$a_k=k/H$, we see that 
$\ln(k L)=\ln(a_k/a_i)$, which is proportional to the total
number of e-folds from the beginning of inflation to the time when the
mode $k$ exits the horizon. This is the large cumulative 
IR correction previously 
alluded to: it gets larger the longer inflation
lasts. Furthermore, the $\ln^2(k\tau)$ accounts for the 
the number of e-folds between horizon exit of
the mode $k$ and, eventually, the end of inflation.
More generally, 
at the generic $n$-th order of perturbation theory, the power spectrum 
is expected to get corrections growing like $\ln^n(kL)$. 

Faced with
these potentially large logarithmic corrections, one may take the road 
of trying to resum them, for example by using the
standard method of the Wilsonian Renormalization Group (RG). 
This technique, however, may not be that  efficient in practice.   
Indeed, the full set of exact RG equations
need to be solved. Approximations inevitably lead to disregarding a
set of diagrams which are not at all subleading in the IR. We refer the reader
to the Appendix for further discussions about this point. 

More relevantly, before attacking the problem of the large IR corrections, 
we should first ask ourselves whether they
are really present in any physical observable which can
be measured. Indeed, one should be suspicious that large effects from
infra-red modes are somehow unphysical, because on general grounds
we expect the structure within our observable patch of the
universe to depend only on its local properties.

What we would like to point out in this short note is
that the large IR corrections appear only in
quantities which are \emph{not directly observable}.
Although their appearance cannot be avoided, they do not themselves have
any particular interest unless one wishes to ask more general 
questions, for instance, what is the probability to find an 
inflaton field homogeneous enough to 
lead to the correct temperature anisotropy in the CMB within patches like ours?
Their presence is an indication that we have computed the answer to
a question which we can never observe. When large IR corrections appear,
we should instead learn how to change the questions we ask
in order to obtain predictions which are genuinely local and apply
within our observable patch.
Some of these considerations already appeared in
Refs.~\cite{lythbox,seery1,seery2} and in the present work
we further elaborate them. 

Let us consider the primordial curvature perturbation computed using the
$\delta N$ formalism
\cite{Starobinsky:1986fx,Sasaki:1995aw,Lyth:2004gb,Lyth:2005fi}, where 
$N$ is the number of e-folds
from an initial flat hypersurface to a final uniform-density hypersurface.
It is convenient to pick the initial time 
to be shortly after Hubble-exit of the scale over which one is smoothing the
relevant physical quantities. The
final slice has to be located before the smoothing scale re-enters the horizon
in such a way that the 
separate universe approach may be employed
\cite{Salopek:1990jq,salopekgrav,Sasaki:1998ug,Wands:2000dp}.
The 
homogenous Friedmann-Robertson-Walker (FRW) equations can be used to calculate
$N$ on large (super-Hubble) scales. 

Assuming for simplicity the presence of a single scalar field $\phi$,
the inflaton,  
the primordial curvature perturbation $\zeta$ is determined by the value of the
scalar field on the initial slice
\begin{equation}
	\zeta(\bx)=\delta N(\phi(\bx))=
		N(\phi(\bx))-N(\overline\phi)=N^{'}\delta\phi(\bx)+\frac{1}{2}N^{''}
		\delta\phi^2(\bx)+\cdots,
	\end{equation}
where 
\begin{equation}
	\delta\phi(\bx)=\phi(\bx)-\overline\phi, \,\,\, N^{'}=
		\left.\frac{dN}{d\phi}\right|_{\overline\phi},\,\,\, 
		N^{''}=\left.\frac{d^2N}{d\phi^2}\right|_{\overline\phi}
		\,\,\,{\rm and}\,\,{\rm so}\,\, {\rm on}.
\end{equation} 
The connected two- and three-point functions of the scalar field
are defined by
\begin{eqnarray}
	\langle\phi_{\bkone} \phi_{\bktwo}\rangle &=&
	P_\phi(k)(2\pi)^3\delta^3(\bkone+\bktwo),\,\nonumber\\
	\langle\phi_{\bkone} \phi_{\bktwo} \phi_{\bkthree}\rangle &=&
	B_\phi(k_1,k_2,k_3)(2\pi)^3
	\delta^3(\bkone+\bktwo+\bktwo),
\end{eqnarray}
where $P_\phi(k)$ is related to the power spectrum $\P_\phi(k)$ by
$\P_\phi(k)=(k^3 P_\phi(k)/2\pi^2)$.
The presence of a nonvanishing bispectrum $B_\phi(k_1,k_2,k_3)$ accounts
(at this order)
for the possibility that the scalar fluctuations are not Gaussian at horizon
exit and, for simplicity, 
we neglect the trispectrum and higher-order connected functions.

We will restrict ourselves from now on to the power spectrum of the
primordial curvature perturbation, $\zeta$,
which is a physical observable within our local patch.
In order to obtain predictions concerning its properties,
all computations should be done within a comoving box
whose present size $\ell$ is not much larger than the present horizon
$H_0^{-1}$, in order that the answer depend only on local properties
of the universe at our position.
In the language of Ref. \cite{lythbox}, this may be called a minimal box.
We shall also be obliged to
consider a superlarge box, whose comoving size $L$ left the
horizon right at the beginning of the inflationary
stage. What we would like to stress in the following is that the appearance of
large logarithmic IR divergences
is a consequence of adopting the superlarge box instead of the minimal one
\cite{lythbox}.

We have said that in order to make predictions for $\zeta$,
we must compute within a box not much larger than $H_0^{-1}$.
However, although we are free \emph{in principle}
to carry out the computation within a box
of any convenient size, in practice we may be restricted by our inability
to compute the relevant correlators.
For example, in chaotic inflation beginning at energy densities close to
the Planck scale, the original inflationary patch will be exponentially larger
than the horizon scale when modes corresponding to the CMB
left the Hubble radius.
In general, our only choice at present is to compute correlation functions
within the entire inflationary region, which constitutes a superlarge box.
However this computation need not have anything to do with the
anisotropy in the CMB temperature: the correlation functions within
the superlarge box are essentially averages of the correlation functions
computed within a horizon-sized box, where the average is taken over all ways
one can fit the small box within the superlarge one.
This correlation function will only successfully predict the CMB anistropy
if conditions in our universe yield correlators which are very close to
those computed by averaging over the superlarge box.

A special simplification occurs if the universe contains only a single field.
In this case, one knows the conditions leading to the end of inflation
and (providing the background field is sufficiently homogeneous)
one therefore always has the choice to compute in a horizon-sized box.
However, where isocurvature fields are present it will generally be true
that these fields assume different values in different regions of the
superlarge box. (This issue is discussed in more detail below.)
In that case, one does not know \emph{a priori} the conditions leading to
the end of inflation in each small box, and the only
alternative in analytic calculations is to
compute in the superlarge box instead.

What is the relationship between the correlation functions in 
differently sized boxes?
As one varies the small box within the superlarge one, there will be
a slow variation in background
quantities such as the Hubble parameter $H$, the slow-roll parameters
$\epsilon$, $\eta$, $\xi$ and so on, together with all other zero-momentum
quantities in the theory\footnote{This sounds contradictory, but
	such quantities are to be thought of as ``zero-momentum'' only within
	the small box. Within the large box they are a superposition of
	modes with wavenumbers in the far infra-red.
	When moving to some larger box, one redefines the spatial average
	of such a quantity as its new ``zero-momentum'' value, and shifts
	some of the infra-red region into the perturbations.}.
The expectation value in the superlarge box
is found by taking averages over these background quantities.
If the background quantities exhibit large variations within the
superlarge box then these averages will develop significant contributions
in the infra-red, associated with wavenumbers between $L^{-1}$ and the
inverse size of the small box.
On the other hand, we could equally well
have obtained the expectation value in the
superlarge box by computing the ensemble average from first principles,
and using the ergodic theorem to connect this with the spatial average.
These two methods of computation must agree.
Therefore, we see that large IR divergences are merely associated with large
fluctuations in the background fields on scales very much larger than the
presently observable region.
These fluctuations may be highly non-gaussian, but because they are
restricted to scales enormously larger than any which are accessible to
experiment, they are of no observational interest.
The presence of such structure on ultra-large scales was
anticipated long ago by Salopek \& Bond \cite{salopekgrav}.

To strengthen the conclusion that the average
$\langle \P_{\ell\zeta} \rangle$ coincides with the power spectrum $\P_\zeta$ 
computed within the superlarge box, we follow Ref. \cite{lythbox} and
consider a box of generic size $M\ll L$ placed within the superlarge
box. We want to show that a quantity such as $\langle P_{M\zeta} \rangle$,
and consequently the average of the power spectrum
$\langle\P_{M\zeta}\rangle$, does not depend upon the scale $M$
and is therefore a reasonable candidate for the expectation value within
a box of size $L$.

The curvature perturbation within the box of size $M$ is given by the
$\delta N$ formula, which yields
\begin{equation}
	\zeta(\bx)=N_M^{'}\delta\phi_M(\bx)+\frac{1}{2}N_M^{''}
	\delta\phi_M^2(\bx)+\cdots,
\end{equation}
where now
\begin{equation}
	\delta\phi_M(\bx)=\phi(\bx)-\overline\phi_M, \,\,\, N_M^{'}=
	\left.\frac{dN}{d\phi}\right|_{\overline\phi_M},\,\,\,
	N_M^{''}=\left.\frac{d^2N}{d\phi^2}\right|_{\overline\phi_M}
	\,\,\,{\rm and}\,\,{\rm so}\,\, {\rm on}.
\end{equation}
We wish to relate this to the same computation performed within the
superlarge box, which entails replacing all quantities evaluated within $M$
by their equivalents evaluated within $L$.
This can be accomplished by using the separate universe
picture to account for variation in the background field, which gives
\begin{eqnarray}
\label{ee}
	\delta\phi_M&=&\left(\overline\phi-\overline\phi_M\right)+
	\delta\phi(\bx),\,\nonumber\\
	N^{'}_M&=& N^{'}_L+N^{''}_L\left(\overline\phi-\overline\phi_M\right)+
	\frac{1}{2}N^{'''}_L\left(\overline\phi-\overline\phi_M\right)^2+\cdots.
\end{eqnarray}
For the sake of simplicity, let us focus on the contributions to the 
two-point correlator coming from the bispectrum (if the scalar field
pertubations are Gaussian at horizon exit, the proof is contained in
Ref.~\cite{lythbox}).
From Ref.~\cite{wandssasaki} we may directly read off the contributions to 
$P_{M\zeta}$ from the bispectrum. They start at one-loop and up to two-loops
they read
\begin{eqnarray}
	P^{\rm 1-loop}_{M\zeta}& \supseteq &
	\frac{1}{(2\pi)^3}\int_{M^{-1}} d^3 q \; N^{'}_M
	N^{''}_M 
	B_{M \phi}\left(k, q, \left|\bkone-\bq\right|\right),\nonumber\\
	P^{\rm 2-loop}_{M\zeta}& \supseteq &
	\frac{1}{(2\pi)^6}\int_{M^{-1}} d^3 q_1 \, d^3 q_2 \; N^{''}_M N^{'''}_M 
	B_{M \phi}\left(\left|\bkone+\bqone\right|, 
	\left|\bqtwo-\bqone\right|,\left|\bktwo-\bqtwo\right|\right)
	P_{M \phi}(q_1),
	\label{running}
\end{eqnarray}
where only momenta larger than $M^{-1}$ are to be considered because the 
box introduces periodic boundary conditions,
and $B_{M\phi}$ is the bispectrum of the inflaton within the $M$-sized
box under consideration. These contributions are
accompanied by terms coming from the power spectrum, as described in
Ref.~\cite{wandssasaki}, which we are ignoring here.
%The tildes over the coefficients $N^{'}_M$ and so on indicate that we are
%using renormalized vertices; in the presence of cubic and higher
%interactions, there are loops that start and finish at the
%same vertex. Instead of including such loops, it has been shown in Ref.
%\cite{wandssasaki} that one can simply replace
%the factors $N^{'}_M$ and so on with their averages 
%\begin{equation}
%	\widetilde{N}^{'}_M =N^{'}_M+\frac{1}{2}N_M^{'''}
%	\langle\delta\phi^2\rangle_M+\cdots,\,\,\,\langle\delta\phi^2\rangle_M=
%	\int_{M^{-1}}\,\frac{dk}{k}\, \P_\phi(k)
%\end{equation}
%and so on. 
%The quantity $\widetilde{N}^{'}_M \widetilde{N}^{''}_M$ can therefore be 
%expanded as
%\begin{equation}
%	\widetilde{N}^{'}_M \widetilde{N}^{''}_M=N^{'}N^{''}+
%	N^{''}N^{'''}(\overline\phi_M-\overline\phi)^2+
%	\frac{1}{2}N^{''}N^{'''}\left((\overline\phi_M-\overline\phi)^2+
%	\langle\delta\phi^2\rangle_M\right) +\cdots.
%\end{equation}

In any $M$-box, the tree-level contribution to $P_{M\zeta}$ is
$(N'_M)^2 P_{M\phi}$. Within the $L$-sized box we can use the
separate universe picture to write
\begin{equation}
	(N'_M)^2 P_{M\phi} = (N'_L)^2 P_{L\phi} +
	\left\{ (N''_L)^2 + N'_L N'''_L + (2\epsilon + \eta) (N'_L)^2 +
	4 \sqrt{2\epsilon} N'_L N''_L \right\}
	(\overline{\phi}_M - \overline{\phi})^2 P_{L\phi} + \cdots ,
	\label{leading-term}
\end{equation}
where a term linear in $(\overline{\phi}_M - \overline{\phi})$ has been
omitted and $\epsilon=(m_{\rm{Pl}}^2/4 \pi) (H'(\phi)/H(\phi))^2$, $\eta= (m_{\rm{Pl}}^2/4 \pi) H''(\phi)/H(\phi)$
are the usual slow-roll parameters. The term proportional to $(N''_L)^2 + N'_L N'''_L$
cancels that part of the running with $M$ which is sourced by the omitted power
spectrum terms in Eq.~(\ref{running}), as described in Ref.~\cite{lythbox}.
Of the remaining terms, $(2\epsilon + \eta) (N'_L)^2$ is of the same
functional form (in derivatives of $N$) as the leading term
$(N'_L)^2 P_{L\phi}$, but is suppressed by extra slow-roll parameters.
This term will cancel contributions from subleading terms
in $P_{L\phi}$ which are not visible when we compute only to leading order.
Therefore this term can be ignored. The interesting term is the one
proportional to $4 \sqrt{2\epsilon} N'_L N''_L$,
which is of the same functional form
in derivatives of $N$ as the one-loop term in Eq.~(\ref{running}).

One must average Eq.~(\ref{leading-term}) over all possible positions
of the $M$-sized box within the large box.
Since $(\overline\phi_M-\overline\phi)$ is simply
$\delta\phi(\bx)$ (within the large box)
smoothed with a top-hat window function, its average is given by
\begin{equation}
	\langle(\overline\phi_M-\overline\phi)^2\rangle=
	\int_{L^{-1}}^{M^{-1}}\,\frac{dk}{k}\, \P_\phi(k).
\end{equation}
Now consider how the one-loop term in Eq.~(\ref{running}) runs with $M$.
Within the $L$-sized box, one can write
(in analogy with Eq.~(\ref{leading-term}))
\begin{equation}
	P_{M\zeta}^{\rm 1-loop} \supseteq
	\frac{1}{(2\pi)^3} \int_{M^{-1}} d^3 q \; \left\{
		N'_L N''_L B_{L\phi} + \mathcal{B}
		(\overline{\phi}_M - \overline{\phi})^2 \right\}
	+ \cdots ,
	\label{one-loop}
\end{equation}
where a linear term has again been discarded,
and $\mathcal{B}$ is an abbreviation for the combination
\begin{equation}
	\mathcal{B} \equiv
		\left\{ \frac{3}{2} N''_L N'''_L + \frac{1}{2} N'_L N''''_L \right\}
			B_{L\phi} +
		\left\{ N'_L N'''_L + (N''_L)^2 \right\} B'_{L\phi} +
		\frac{1}{2} N'_L N''_L B''_{L\phi}
	\label{b-def} .
\end{equation}
In this equation,
$B'_{L\phi}$ and $B''_{L\phi}$ are, respectively, the first and second
derivatives of the inflaton bispectrum within the $L$-sized box with
respect to the background field, and all the bispectra are evaluated with
arguments $k$, $q$ and $|{\bf k}_1 - {\bf q}|$. The leading term in
Eq.~(\ref{one-loop}) is proportional to $N'_L N''_L B_{L\phi}$.
The running of this term comes entirely from differentiating the integral
with respect to its limit. Using the method of Boubekeur \& Lyth
\cite{Boubekeur:2005fj} to estimate the integral, and the squeezed
bispectrum computed by Allen \emph{et al.} \cite{Allen:2005ye}
(see also Ref.~\cite{Cheung:2007sv}),
it follows that this running exactly cancels the running from the
term proportional to $N'_L N''_L$ in Eq.~(\ref{leading-term}).

The remaining terms in Eq.~(\ref{one-loop}), which are described by
$\mathcal{B}$, acquire running from both
the lower limit of the integral, and the background
expectation value $\langle (\overline{\phi}_M - \overline{\phi})^2 \rangle$.
They are proportional to three copies of $P_{\phi}$ and therefore
are formally comparable to two-loop corrections. In a full calculation, one
should include all two-loop terms and show that the running
of the $\mathcal{B}$ terms correctly cancels that of the two-loop
contributions. In this paper we focus
on the two-loop term written in Eq.~(\ref{running}) and show that its
running cancels with the term of the same functional form
(in derivatives of $N$) in $\mathcal{B}$.
The running imparted to this term by $\langle
(\overline{\phi}_M - \overline{\phi})^2 \rangle$ is given by
\begin{equation}
	\label{onelooprunning}
	\frac{d}{d\ln M}\langle P^{\rm 1-loop}_{M\zeta}\rangle \supseteq
	\frac{3}{2} \frac{1}{(2\pi)^3}N^{''}_L N^{'''}_L \P_{L \phi}(M^{-1})
	\int_{M^{-1}} d^3 q \;
	B_{L \phi}\left(k, q, \left|\bkone-\bq\right|\right) .
\end{equation}
This can be thought of as a renomalization of the classical value of the
scalar field as seen inside the small box. 

On the other hand,
the $M$-dependence of $P^{\rm 2-loop}_{M\zeta}$ comes by
differentiating the integral with respect to its limit 
\begin{equation}
	\label{aa}
	\frac{d}{d\ln M}\langle P^{\rm 2-loop}_{M\zeta}\rangle \supseteq
	- \frac{1}{(2\pi)^3}N^{''}_L N^{'''}_L \P_{L \phi}(M^{-1})
	\int_{M^{-1}} d^3 q \; 
	B_{L \phi}\left(k, q, \left|\bkone-\bq\right|\right),
\end{equation}
where we have assumed that the relevant physical momenta are much larger than
$M^{-1}$.  The contribution~(\ref{aa}) cancels two thirds of the running in
Eq.~(\ref{onelooprunning}). The remaining part is associated with
``dressing'' of the $\delta N$ coefficients, as described in
Ref.~\cite{wandssasaki}. It is cancelled by a bispectrum term coming
from the $N''_L N'''_L$ term in the $\delta N$ expansion, leaving zero running
overall.
A similar computation may be 
performed including the trispectrum. 

The computation above confirms that $\langle \P_{M\zeta} \rangle$ does not
depend on the size of the box $M$, as expected on general grounds,
and that it coincides with the power spectrum $\P_\zeta$ computed in the
superlarge box: large IR divergences inevitably appear because what we are
computing is in fact $\langle \P_{\ell\zeta} \rangle$, that is the
power spectrum on the superlarge box. 

One can choose a small box to minimize the two-loop contribution, but this
makes the one-loop contribution sensitive to the
largest scale $L$.
For instance, in slow-roll models
of inflation $B_\phi={\cal O}(\epsilon^{1/2})P_\phi^2$. It can be shown that
\begin{equation}
	\langle \P^{\rm 1-loop}_{\ell\zeta} \rangle={\cal O}(\epsilon^2)\,
	\P_\zeta^2\ln(L/M) ,
\end{equation}
which becomes significant for $M \ll L$.
Alternatively, one can try
to reduce the one-loop contribution going to the superlarge box, paying the
price of a large two-loop contribution. 
The point though is that the large IR divergences show up in the average
power spectrum and, as we have advocated before, this is not a quantity
of physical interest. It only provides the level of uncertainty inherent in the
theoretical predictions. 

If one insists in adopting a superlarge box, it should be kept in mind
therefore that the location of our box
may be untypical and one should quantify how likely it is that the correlators
averaged over the superlarge box coincide with the correlators in
our observable universe.
To deal with this problem, one can try to use the approach of stochatic
inflation.
This embodies the idea that the IR part of the scalar field may be
considered as a classical space-dependent
stochastic field satisfying a local Langevin-like equation \cite{starob}. 
The stochastic noise terms arise from the quantum fluctuations which become
classical at horizon crossing and then contribute to the background. 
One can derive
a Fokker-Planck equation describing how the probability of scalar field values
at a given spatial point evolves with time. One should split the computations
into two steps. First, the distribution of the values of the
scalar field inside the superlarge box should be estimated.
This will allow one to compute the probability
$P(\overline{\phi}_{60}\left.\right|t_{60})$
that the starting initial condition for the background field is given by
$\overline\phi= \overline{\phi}_{60}$ in our local observable patch at time
$t_{60}$, {\it i.e.} 
when there remain 60 or so e-folds till the end of inflation in that patch.
The distribution of the scalar field in the superlage box, due to the IR
divergences, will be far from being Gaussian. 
However, we do expect that in our local patch initial conditions 
are relatively uniform. From the point of view of our local universe
the IR properties of the superlarge box only enter to provide the probability 
$P(\overline{\phi}_{60}\left.\right|t_{60})$ that 
the value $\overline{\phi}_{60}$ is achieved in our patch. At later times the
field value at a given point has a distribution given by
\begin{equation}
	P(\overline{\phi}\left.\right|t)=\int d\overline{\phi}_{60}
	P\left(\overline{\phi}\left.\right|t; t_{60},\overline{\phi}_{60}
	\right) P(\overline{\phi}_{60}\left.\right|t_{60}),
\end{equation}
Furthermore, if the scalar field is homogeneously distributed
in our local patch \cite{salopekgrav}, then
	$P\left(\overline{\phi}\left.\right|t_{60}; 
	t_{60},\overline{\phi}_{60}
	\right)=\delta\left(\overline\phi-\overline\phi_{60}\right)$. 
The probability $P(\overline{\phi}\left.\right|t)$ 
is generically  highly non-Gaussian, because the  initial conditions are not
Gaussian distributed as large IR corrections are present in the superlarge
box 
\cite{salopekgrav,mmol}.
However, if we impose the observable constraint in order to select those
initial parameters which lead to CMB anisotropies
consistent with observations, a severe selection is imposed on initial
conditions and we expect a probability close to Gaussian. 
In the stochastic approach, therefore, one trades the uncertainty in
the prediction coming from the large IR contributions in the superlarge box for
the uncertainty inherent in having a probability distribution
for the background quantities. It would be interesting to determine such a 
probability distribution to evaluate the uncertainty in the theoretical 
predictions for our local universe.

\acknowledgments
\noindent
The authors are grateful to David Lyth for useful conversations and comments. 
This research was supported in part by the European Community's Research
Training Networks under contracts MRTN-CT-2004-503369 and MRTN-CT-2006-035505.

\appendix

\section{The in--in formalism for
non-equilibrium quantum field theory}

\noindent
In this Section  we    briefly present    some of the  basic
features of the  non-equilibrium quantum field theory based on the
in-in, also dubbed Schwinger-Keldysh,  formulation \cite{sk}.
The interested reader is referred to the excellent review by Chou
{\it et al.} \cite{chou} for a more exhaustive discussion.

Since  we
need the temporal evolution of quantum correlators with definite
initial conditions and not
simply the transition amplitude of particle reactions,
the ordinary equilibrium quantum field theory at finite temperature
is not the appropriate tool.
The most appropriate extension of the field theory
to deal with nonequilibrium phenomena amounts to generalizing 
the time contour of
integration to a closed-time path. More precisely, the time integration
contour is deformed to run from $-\infty$ to $+\infty$ and back to
$-\infty$.

 The CTP formalism (often  dubbed as in-in formalism) is a powerful
Green's function
formulation for describing non-equilibrium phenomena in field theory.  It
allows to describe phase-transition phenomena and to obtain a
self-consistent set of quantum Boltzmann equations.
The formalism yields various quantum averages of
operators evaluated in the in-state without specifying the out-state.
On the contrary, the ordinary quantum field theory (often dubbed as in-out
formalism) yields quantum averages of the operators evaluated
with an in-state at one end and an out-state at the other.

\begin{figure}
\scalebox{0.5}{\includegraphics{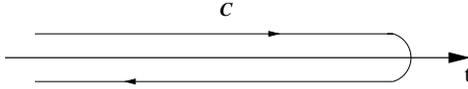}} 
\caption{The appropriate time contour $C$.}
\label{fig0}
\end{figure}

For example, because of the time contour deformation, the partition function
in the in-in formalism for a real scalar field is defined to be
\begin{eqnarray}
\label{path}
Z\left[ J\right] &=& {\rm Tr}\:
\left[ T\left( {\rm exp}\left[i\:\int_C\:J\phi\right]\right)\rho\right]\nonumber\\
&=& {\rm Tr}\:\left[ T_{+}\left( {\rm exp}\left[ i\:\int\:
J_{+}\phi_{+}\right]\right)
\right.
\nonumber\\
&\times&\left.  T_{-}\left( {\rm exp}\left[
      -i\:\int\:J_{-}\phi_{-}\right]\right) \rho\right],
\end{eqnarray}
where $C$ in the integral denotes that the time
integration contour runs from minus infinity to plus infinity
and then back to minus infinity again, see Fig. \ref{fig0}. The symbol $\rho$
represents the initial density matrix and the fields are in
the Heisenberg picture  and  defined on this closed time contour.
Sometimes it is more usful to work with other field variables,
$\phi_c=1/2(\phi_+ + \phi_{-})$ and $\phi_\Delta=(\phi_+ - \phi_{-})$. In such a case one has to
properly redefine the sources as $J_c=1/2(J_+ + J_{-})$ and $J_\Delta=(J_+ - J_{-})$. To identify
the physical degrees of freedom, the normalization of the generating functional
$\left.Z\left[J_\Delta,J_c\right]\right|_{J_\Delta=0}=1$ has to be imposed.

We must now identify field
variables with arguments on the positive or negative directional
branches of the time path. This doubling of field variables leads to
six  different real-time propagators on the contour \cite{chou}.  These six
propagators are not independent, but using all of them simplifies the
notation.
For a generic bosonic neutral  scalar field $\phi$ they are defined as
\begin{eqnarray}
\label{def1}
G^{-+}\left(x, y\right)&=&
i\langle
\phi(x)\phi (y)\rangle,\nonumber\\
G^{+-}\left(x, y\right)&=&i\langle
\phi (y)\phi(x)\rangle,\nonumber\\
G^{++}\left(x, y\right)&=&
 G^{-+}(x,y)\theta(x,y)+
G^{+-}(x,y)\theta(y,x),\nonumber\\
G^{--}\left(x, y\right)&=&
 G^{+-}(x,y)\theta(y,x)+
 G^{-+}(x,y)\theta(x,y), \nonumber\\
G^R(x,y)&=&G^{++}(x,y)-G^{+-}(x,y)=\left(G^{-+}(x,y)-G^{+-}(x,y)\right)\theta(x_0-y_0),\nonumber\\
G^A(x,y)&=&G^{++}(x,y)-G^{-+}(x,y)=\left(G^{+-}(x,y)-G^{-+}(x,y)\right)\theta(y_0-x_0),
\end{eqnarray}
where the last two Green functions are the retarded and advanced
Green functions respectively and $\theta(x,y)=\theta(x_0-y_0)$
is the step function. When computing a loop diagram, one has to assign to the  interaction points
a plus or a minus sign in all possible manners and sum all the possible
diagrams, taking into account that vertices which  a minus sign
has been assigned to must be multiplied by $-1$.

In the basis of the fields $\phi_c$ and $\phi_\Delta$, one may define
the Green functions

\begin{eqnarray}
\langle\phi_c(x)\phi_c(y)\rangle &\equiv & G^c(x,y)=-\frac{i}{2}\left(G^{-+}(x,y)+G^{+-}(x,y)\right),
\nonumber\\
\langle\phi_c(x)\phi_\Delta(y)\rangle &=& -i \,G^R(x,y),
\end{eqnarray}
while the Green function $\langle\phi_\Delta(x)\phi_\Delta(y)\rangle$ vanishes identically because of the identity $G^R+G^A=G^{+-}+G^{-+}$.

For equilibrium phenomena, the brackets $\langle \cdots\rangle$
imply a thermodynamic average over all the possible states
of the system. While for homogeneous systems in equilibrium, the
Green
functions
depend only upon the difference of their arguments $(x,y)=(x-y)$
and there is no dependence upon $(x+y)$,
for systems out of equilibrium, the
definition (\ref{def1})
has a different meaning. The concept of thermodynamic averaging
is now ill-defined. Instead, the bracket means the need to
average over all the available states of the system for the non-equilibrium
distributions. Furthermore, the arguments of the Green functions
$(x,y)$ are {\it not} usually given as the difference $(x-y)$.
For example, non-equilibrium could be caused
by transients which make the Green functions
depend upon $(x_0,y_0)$ rather than $(x_0-y_0)$.

\begin{figure}
\scalebox{0.5}{\includegraphics{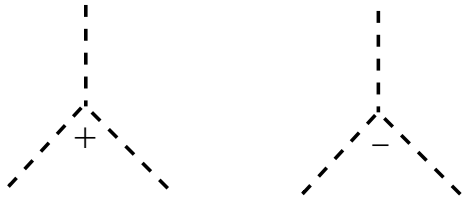}} 
\caption{The cubic vertices with the proper signs $+$ and $-$.}
\label{fig1}
\end{figure}

The Lagrangian we consider is the one for a massless (at the tree level) scalar field
with a cubic self-interaction. This field is not necessarily the scalar field
driving inflation.

\begin{equation}
{\cal L}[\phi_+]-{\cal L}[\phi_-]=\sqrt{-g}\left(\frac{1}{2}g^{\sigma\rho}(\partial_\sigma\phi_+\partial_\rho\phi_+
-\partial_\sigma\phi_-\partial_\rho\phi_-)
-\frac{\mu}{3!}(\phi_{+}^3-\phi_{-}^3)
\right).
\end{equation}
We are using the conformal  metric $g_{\sigma\rho}=a^2(\tau)\,{\rm diag}(1,-1,-1,-1)$, where
$a(\tau)=-1/H\tau$ ($\tau<0$ is the conformal time)
 is the scale factor during the de Sitter stage characterized by
a Hubble rate $H$. The interaction term is cubic in the scalar field, 
see Fig. \ref{fig1},  and $\mu$ has
therefore the dimensions of a mass.

\begin{figure}
\scalebox{0.5}{\includegraphics{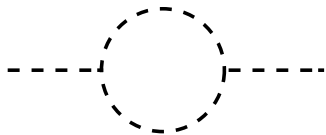}} 
\caption{The diagram contributing to the one-loop two-point 
correlation function. To each vertex the signs $+$ 
and $-$ should be attached to each vertex according to the
correlator one wishes to compute.}
\label{fig2}
\end{figure}

Since we will be interested in the following in the infra-red 
divergences, we provide here the
expressions for the Fourier transformed of the tree level expressions 
$g^c$ and $g^R$
of the
two-point correlation functions $G^c$ and $G^R$ respectively in
momentum space for wavelengths larger than the horizon

\begin{equation}
g_{\vec{k}}^c(\tau_1,\tau_2)\simeq\frac{H^2}{2k^3},\,\,\,
g_{\vec{k}}^{R}(\tau_1,\tau_2)\simeq\frac{H^2}{3k^3}\theta(\tau_1-\tau_2)\left[k^3(\tau_1^3-
\tau_2^3)\right],\,\,\,\,\,(k\tau_1,k\tau_2\ll 1).
\label{tree}
\end{equation}
As an illustrative example of the infra-red divergences, we compute the 
two-point correlation function $G_{\vec{k}}^c(\tau,\tau')$.
The relevant diagram is given in Fig. \ref{fig2} 
which represents
the sum of a  set 
of diagrams whose relative sign is dictated by the appropriate
sign of the cubic vertex.

A straightforward calculation which makes use of the expressions 
(\ref{tree}), shows that the infra-red contribution to
$G_{\vec{k}}^c(\tau,\tau')$
at one-loop is given by

\begin{eqnarray}
\label{twopoint}
G_{\vec{k}}^{c\,{\rm 1-loop}}(\tau,\tau')&\simeq& \mu^2\int^\tau d\tau_1 \,a^4(\tau_1)\int^{\tau_1}  d\tau_2
\, a^4(\tau_2)\,g^R_{\vec{k}}(\tau,\tau_1)\,g_{\vec{k}}^c(\tau',\tau_2)
\int\frac{d^3p}{(2\pi)^3}\,g^R_{\vec{k}+\vec{p}}(\tau_1,\tau_2)
g_{\vec{p}}^{c}(\tau_1,\tau_2)\nonumber\\
&+&\mu^2\int^{\tau'} d\tau_1 \,a^4(\tau_1)\int^{\tau_1}  d\tau_2
\, a^4(\tau_2)\, g^R_{\vec{k}}(\tau',\tau_1)\, g_{\vec{k}}^c(\tau,\tau_2)\int\frac{d^3p}{(2\pi)^3}\, g^R_{\vec{k}+\vec{p}}(\tau_1,\tau_2)\,
g_{\vec{p}}^c(\tau_1,\tau_2)\nonumber\\
&+&\frac{1}{2}\,
\mu^2\int^\tau d\tau_1 \, a^4(\tau_1)\int^{\tau'}  d\tau_2
\, a^4(\tau_2)\,g^R_{\vec{k}}(\tau,\tau_1)\,
\int\frac{d^3p}{(2\pi)^3}\,g_{\vec{p}}^c(\tau_1,\tau_2)\, g^R_{\vec{k}}(\tau',\tau_2)
\, g_{\vec{k}+\vec{p}}^c(\tau_1,\tau_2)\nonumber\\
&=&\frac{1}{9}\frac{\mu^2 g_{\vec{k}}^c(\tau,\tau')}{(2\pi)^2 H^2}\left(\int^\tau 
d\tau_1\int^{\tau_1}d\tau_2 \frac{(\tau^3-\tau_1^3)(\tau_1^3-\tau_2^3)}{(\tau_1\tau_2)^4}
+\int^{\tau'} d\tau_1\int^{\tau_1}d\tau_2
\frac{({\tau'}^3-\tau_1^3)(\tau_1^3-\tau_2^3)}{(\tau_1\tau_2)^4}\right.\nonumber\\
&+&\left.\int^\tau d\tau_1\int^{\tau'}d\tau_2
\frac{({\tau}^3-\tau_1^3)({\tau'}^3-\tau_2^3)}{(\tau_1\tau_2)^4}\right)\left(
\int^k_{L^{-1}}
\frac{dp}{p}\right)
\nonumber\\
&\simeq&
\frac{1}{18}\frac{\mu^2 g_{\vec{k}}^c(\tau,\tau')}{(2\pi)^2 H^2}
\ln(k L)\ln^2(k^2\tau\tau'),\nonumber\\
&&
\end{eqnarray}
where $L^{-1}$ is the comoving infra-red cut-off which 
we may set to be equal to
$a_i H$, $a_i$ being the 
value of the scale factor at the beginning of inflation (the
upper limit of integration over the momenta $p$ is dictated by the fact that
the expression is valid for  momenta $p\lsim k$).
Notice that the power spectrum $\P_\phi$ of the perturbations
of the scalar field $\phi$ is directly related to
the correlation function $G_{\vec{k}}^c(\tau,\tau')$ computed at equal
times
\begin{figure}
\scalebox{0.5}{\includegraphics{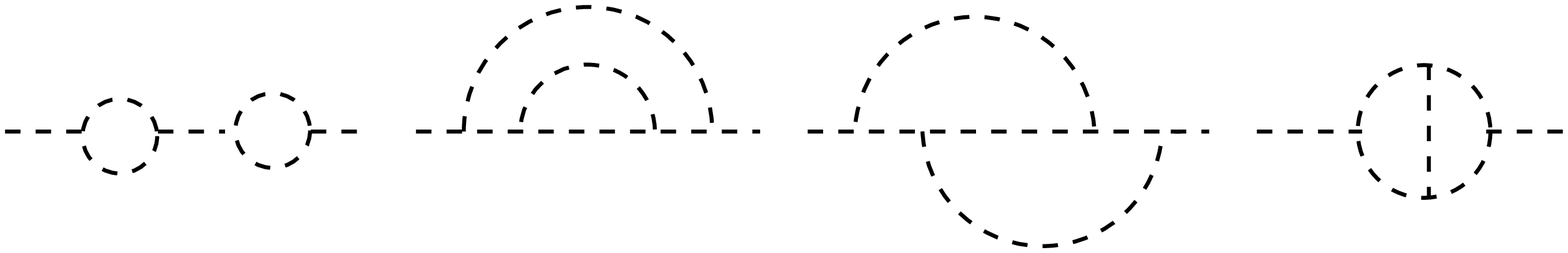}} 
\caption{The diagrams contributing to the two-loop two-point 
correlation function. To each vertex the signs $+$ 
and $-$ should be attached to each vertex according to the
correlator one wishes to compute.}
\label{fig3}
\end{figure}
\noindent
\begin{equation}
\P_\phi(k)=\frac{k^3}{2\pi^2}G_{\vec{k}}^c(\tau,\tau).
\end{equation}
A similar computation leads to the 
one loop correction to the retarded Green function
in the infra-red

\begin{equation}
G^{R\,{\rm 1-loop}}_{\vec{k}}(\tau,\tau')\simeq 
\frac{2}{9}\frac{\mu^2 g^R_{\vec{k}}(\tau,\tau')}{(2\pi)^2H^2}\ln(k L)
\ln^2(\tau/\tau').
\end{equation}
At two-loop, one can easily calculate the  IR contribution
of the first three graphs in Fig. \ref{fig3} obtaining

\begin{equation}
\label{twoloop}
G_{\vec{k}}^{R\,{\rm 2-loop}}(\tau,\tau')\simeq
\frac{2\mu^4 }{81(2\pi)^4H^4}g_{\vec{k}}^R(\tau,\tau')
\ln^2(k L)\ln^4\tau/\tau'). 
\end{equation}
It is not difficult to show that the same 
kind of topology of diagrams appearing at higher loops leads to 
the resummation

\begin{equation}
G_{\vec{k}}^{R\,{\rm res}}(\tau,\tau')\simeq g_{\vec{k}}^R(\tau,\tau'){\rm exp}
\left(\frac{2}{9}\frac{\mu^2}{(2\pi)^2H^2}
\ln(k L)\ln^2(\tau/\tau')\right).
\end{equation}
However, diagrams like the forth one in Fig. (\ref{fig3}) are not
automatically included in this resummation. This is unfortunate, because they
are not subleading. 
This is reminiscent of what  happens
in a $\lambda\phi^4$ theory where, performing   a Hartree approximation,
one is able to resum the all class of ``daisy'' diagrams;  however, unless
an $O(N)$ symmetry is invoked for large $N$, one has no guarantee that the
omitted diagrams are unimportant. 

\begin{figure}
\scalebox{0.5}{\includegraphics{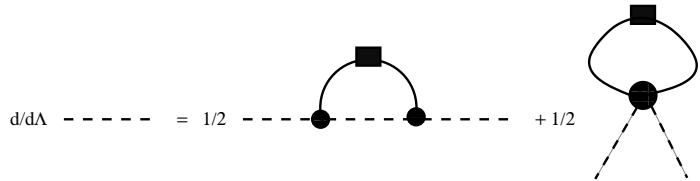}} 
\caption{The RG equation for the two-point correlation function.}
\label{rge}
\end{figure}

In general, one can hope to resum the large
leading logarithms using the
standard method of the Wilsonian 
Renormalization Group  (RG) techniques developed in quantum field
theory
and in statistical physics \cite{book}. 
These techniques have been recently applied to resum cosmological
perturbations of a collisionless fluid \cite{MP}. 
The starting point of this formulation of  the Renormalization Group  is 
a modification
of the generic two-point correlation function as follows

\begin{equation}
G_{\vec{p}}(\tau_1,\tau_2)\rightarrow G^\Lambda_{\vec{p}}(\tau_1,\tau_2)=
G_{\vec{p}}(\tau_1,\tau_2)\theta(p,\Lambda),
\end{equation}
where $\theta(p,\Lambda)$ is a high-pass filtering function, which equals unity
if $p\gg \Lambda$ and zero if $p\ll\Lambda$.
The modified generating functional
$Z_\Lambda[J]$ can be obtained from the expression (\ref{path}) and  
describes
a fictitious universe in which all fluctuations with momenta smaller 
than $\Lambda$
are damped out. In the limit $\Lambda\rightarrow L^{-1}$ 
all fluctuations are included
and we recover the physical situation. Decreasing the cut off, 
the linear and the nonlinear effects of smaller and smaller fluctuations 
are gradually
taken into account. This process is described  by the RG equation 
obtained  by taking
the $\Lambda$ derivative of $Z_\Lambda$

\begin{equation}
\Lambda\frac{\partial }{\partial\Lambda}\left. 
Z_\Lambda[J_c, J_\Delta]\right|_{J_c=0}=-
\frac{i}{2}\,{\rm Tr}\,\frac{\delta}{\delta J_\Delta}
\cdot\Lambda\frac{\partial }{\partial\Lambda}(g^\Lambda)^{-1}
\cdot\frac{\delta}{\delta J_\Delta}
\left. Z_\Lambda[J_c,J_\Delta]\right|_{J_c=0},
\end{equation}
where the trace stands for integration over momenta and conformal time. 
Differentiating with
respect to the source $J_\Delta$ the generating functional 
$W_\Lambda=-i\ln Z_\Lambda$, one can easily find the RG equations 
for the connected Green
functions. The exact RG equation at any order in 
perturbation theory
 for the two-point correlation function is
symbolically given in Fig. \ref{rge} where the 
dots indicated the fully renormalized
vertex, the propagators are meant to be fully renormalized and the propagators
with the square represents the RG kernel  proportional to 
$\Lambda\frac{\partial }{\partial\Lambda}(g^\Lambda)^{-1}$.

RG methods are particularly suited to physical situations in which there 
is a separation
between the scale where one is supposed to control the fundamental theory 
(in our case on subhorizon scales at the beginning of  inflation) and the
scale at which the measurements are actually made 
(on superhorizon scales at the end of inflation). 
One can fix the initial conditions for a generic two-point correlation function
$G_{\vec{p}}(\tau_1,\tau_2)$ at the UV scale $\Lambda_{UV}=(a_p H)$. In this
way, the correlator receives contributions only from those modes which
are still inside the horizon, being $a_p$ the scale factor at which
the mode $p$ exits the horizon, $a_p=(p/H)$. In this range of modes
the correlator is supposed to be known and equal to the one given in
Minkowski spacetime once the renormalization procedure
is accounted for and all the parameetrs of the theory have been properly 
renormalized. Notice that, when computing the correlator
for a given external momentum $k$, the UV cut off on the internal momenta 
will be chosen to be $(a_k H)$, since 
one is  interested in the effect of the internal superhorizon modes
which are well out of the horizon when the mode $k$ exits the horizon. 

The cut-off scale $\Lambda$ is then 
decreased in such a way that more and more momenta corresponding to
wavelengths outside the horizon are included. The cut off scale  $\Lambda$
is decreased till it reaches the smallest momentum which is excited during the
de Sitter stage, that is up to  $\Lambda_{IR}=\Lambda_{UV}(a_i/a_p)=H$, where
$a_i$ is the scale factor at the beginning of inflation. THis corresponds
to the comoving IR cut-off given by $L^{-1}$.  
Starting from the fundamental scale, the RG flow
describes the gradual inclusion of fluctuations at scales closer and 
closer to the one relevant
to measurements. The key point is that the 
new fluctuations which are included at each intermeadiate
step feel an effective theory which 
has been dressed by the fluctuations already included. 

The problem with this approach, however, is that the full set of RG
equations for  $G^c$ and $G^R$ has to be solved as  both two-point
functions are equally renormalized in the IR.  Solving 
an approximate version of the RG equations
corresponds to resum only a given
class of diagrams, thus neglecting others which in most cases are not
subleading at all. For instance, in the example developed in this
Appendix of a cubic theory, neglecting the vertex renormalization 
and   taking free propagators on the
right-hand side of the RG equations amounts to resum that class of diagrams
which at two-loop is represented by the first three graphs 
of Fig. \ref{fig3}. The inclusion of equally important diagrams in the
IR amounts to solving the full set of 
RG equations with less and less approximations, 
making the approach very complicated.

\end{document}